\documentclass[12pt]{article}
\pdfoutput=1
\usepackage{latexsym,cite}
\usepackage{amsmath}
\usepackage{amssymb}
\usepackage{amsthm}
\usepackage{subcaption}
\usepackage{mathtools}
\usepackage{braket}

\usepackage[top = 1. in,bottom = 1 in, left = 1 in, right=0.9 in]{geometry}

\newcommand{\arXiv}[1]{\href{http://www.arXiv.org/abs/#1}{arXiv:#1}}
\newcommand{\doi}[2]{\href{http://dx.doi.org/#2}{#1}}
\usepackage[colorlinks=true, linkcolor=blue, bookmarks=true]{hyperref}

\usepackage{tikz}
\usepackage{xcolor}
\definecolor{hookone}{RGB}{255,190,190}
\definecolor{hooktwo}{RGB}{190,220,255}
\definecolor{hookthree}{RGB}{190,245,190}
\definecolor{hookfour}{RGB}{255,220,160}

\usepackage{array,booktabs,longtable}

\newcommand{\youngdiagram}[1]{%
  \begin{tikzpicture}[scale=0.45]
    \foreach \y [count=\row from 0] in {#1} {
      \foreach \x in {1,...,\y} {
        \draw (\x-1,-\row) rectangle (\x,-\row-1);
      }
    }
  \end{tikzpicture}%
}

\allowdisplaybreaks

\makeatletter
\renewcommand\section{\@startsection {section}{1}{\z@}%
                  {-3.5ex \@plus -1ex \@minus -.2ex}
                  {2.3ex \@plus.2ex}%
                  {\normalfont\large\bfseries}}
\renewcommand\subsection{\@startsection{subsection}{2}{\z@}%
                   {-3.25ex\@plus -1ex \@minus -.2ex}%
                   {1.5ex \@plus .2ex}%
                   {\normalfont\bfseries}}
\makeatother

\newcommand{\beq}{\begin{equation}}
\newcommand{\eeq}{\end{equation}}
\newcommand{\del}{\partial}

\newcommand{\al}{\alpha}
\newcommand{\ab}{\bar\alpha}
\newcommand{\te}{\theta}
\newcommand{\ze}{\zeta}
\newcommand{\Sz}{Szeg\H{o}}

\newcommand{\ad}{a^\dagger}

\newcommand{\psmax}[2]{\psi_{\scriptscriptstyle{#1, #2}}}

\begin{document}
\begin{titlepage}
\begin{flushright}
\phantom{arXiv:yymm.nnnn}
\end{flushright}
\begin{center}
{\huge\bf The integer spectrum of a\vspace{2mm}\\ superintegrable quantum field theory}
\vskip 12mm
{\large Marine De Clerck}
\vskip 3mm
{\it DAMTP, University of Cambridge, Cambridge, United Kingdom}
\vskip 7mm
{\large Oleg Evnin}
\vskip 3mm
{\it  High Energy Physics Research Unit, Faculty of Science, \\Chulalongkorn University,
Bangkok, Thailand\vspace{1mm} \\ Theoretische Natuurkunde, Vrije Universiteit Brussel (VUB) \\
\&\,\,
International Solvay Institutes, Brussels, Belgium}
\vskip 7mm
{\small {\tt md989@cam.ac.uk, oleg.evnin@gmail.com}}
\vskip 35mm
{\bf ABSTRACT}\vspace{3mm}
\end{center}

In our previous article {\it CMP 407 (2026) 131}, we explored the rich algebraic structures arising from a specific quantum many-body Hamiltonian quartic in the bosonic creation-annihilation operators. This Hamiltonian originated from the works of G\'erard and Grellier on classical turbulent energy transfer in a Lax-integrable setting, and its quantum version respects hierarchies of polynomial conservation laws all of which possess purely integer spectra, as does the Hamiltonian itself. The latter feature, known from numerical studies, has previously remained unproved.
Here,  we present an elementary solution of the quantum Gérard-Grellier Hamiltonian. The eigenstates are labelled by partitions and the corresponding eigenvalues admit simple explicit expressions in terms of the principal hook lengths. The key ingredient in this construction is a specific basis we introduce, inspired by the classical multisoliton coherent states, in which the Hamiltonian is upper-triangular. The expressions for the eigenvalues then immediately follow, while the full eigenbasis can be recovered via a Gram-Schmidt procedure.

\vfill

\end{titlepage}

\section{Introduction}

Turbulence and integrability are two landmark topics in studies of classical field dynamics.
G\'erard and Grellier succeeded in bringing these topics together in a fruitful manner \cite{GG} by introducing the deceptively simple equation
\beq\label{cuSz}
i\,\del_t u =\Pi(|u|^2u)
\eeq
for a function $u(t,\te)$ on the spatial unit circle $\te\in[0,2\pi)$ that is naturally embedded in the complex plane $\ze$ by writing $\ze=e^{i\te}$.
Here, $\Pi$ is the \Sz{} projector:
\beq
\Pi \left( \sum_{k=-\infty}^\infty\al_k e^{i k\te} \right) =\sum_{k=0}^\infty\al_k e^{i k\te},
\eeq
and one naturally looks for solutions satisfying $\Pi( u)=u$ so that
\beq
u(t,\te)=\sum_{k=0}^\infty \al_k(t)\,e^{ik\te},\quad\mbox{or}\quad u(t,\zeta)=\sum_{k=0}^\infty \al_k(t)\,\zeta^k.
\eeq
This construction justified the name {\it `cubic \Sz{} equation'} used for (\ref{cuSz}) in \cite{GG}. Pictorially, one is talking about the transfer of energy between the normal modes of interacting waves on a unit circle that are only allowed to move in one of the two directions (a behavior physically familiar from chiral edge modes).

It turned out that equation (\ref{cuSz}) is both Lax-integrable and displays prominent turbulent behaviors (transfer of excitation towards progressively shorter spatial scales); a review can be found in \cite{GGrev}. One of the most apparent properties of the corresponding dynamics is the presence of self-consistent multisoliton solutions realized by the very simple expression
\beq\label{classinv}
u(t,\zeta)=c(t)+\sum_{r=1}^R\frac{c_r(t)}{1-p_r(t)\zeta}
\eeq
for any assignment of $R$, interpreted as the number of solitons. (The first term $c(t)$ may be optionally omitted.) The solitons are thus realized as poles of a meromorphic function moving in the complex plane, with positions determined by $p_r(t)$. The integrability structure inherent to (\ref{cuSz}) is so constraining that it allows, remarkably, for writing down an explicit general solution for prescribed initial data \cite{GGsol}. These seminal developments have given rise to substantial further investigations and have ultimately led to similar solutions of a few other related integrable systems \cite{GK,explicit,CMNLS,halfwavemap}.

Equation (\ref{cuSz}) can be naturally derived from the Hamiltonian $\int d\te\, |u|^4$, which is written in terms of the mode expansion as $\sum_{n+m=k+l} \ab_n\ab_m \al_k \al_l$. The latter form is structurally a special case
of {\it resonant Hamiltonian systems} commonly arising as weakly nonlinear approximations to dynamics in resonant domains \cite{resrev}.
Any Hamiltonian system can be straightforwardly quantized, and in the specific case we are considering, the resulting {\it quantum G\'erard-Grellier (GG) Hamiltonian}
is given by
\begin{equation}\label{eq: HGG}
H\equiv \frac12\sum^{i+j=k+l}_{i,j,k,l\ge 0} a^\dagger_i a^\dagger_j a_k a_l,
\end{equation}
where $a^\dagger_k$ and $a_k$ labelled by an integer $k\ge 0$ are conventional bosonic creation-annihilation operators satisfying the standard commutation relations $[a_k, a_l^\dagger]=\delta_{kl}$.
The Hamiltonian commutes with the particle number and momentum operators
\begin{equation}\label{eq: NMq}
\hat N=\sum_{k=0}^\infty a^\dagger_k a_k,\qquad \hat M=\sum_{k=1}^\infty k  \,a^\dagger_k a_k \, .
\end{equation}
(We use hats here to distinguish the quantum operators $\hat N$ and $\hat M$ from their integer eigenvalues $N$ and $M$.)
In the standard Fock representation, basis states are labelled by integer partitions $\lambda \in \mathcal{P}_N(M)$ of $M$ units into at most $N$ parts, with the occupation number of mode $i$ in the state $|\lambda\rangle$,  denoted $\eta_i(\lambda)$, given by the multiplicity of part $i$ in the integer partition $\lambda$ (and mode 0 filled in with an appropriate number of particles $\eta_0$ to ensure that the total number of particles is $N$). Explicitly,
\begin{align}|\lambda \rangle = |\eta_0,\eta_1,\cdots\rangle\equiv \left(\prod_{k=0}^\infty \frac{(a^\dagger_k)^{\eta_k}}{\sqrt{\eta_k!}}\right)|0,0,\cdots\rangle, \\
\text{with} \quad a_k |0,0,\cdots\rangle=0,\quad \hat \eta_ k |\eta_0,\eta_1,\cdots\rangle=\eta_k|\eta_0,\eta_1,\cdots\rangle \, \nonumber
\label{eq: Fock}
\end{align}
with $\hat \eta_ k\equiv a^\dagger_k a_k$. The matrix elements of $H$ between Fock states with different values of $N = \sum_k\eta_k$ and $M =\sum_k k\,\eta_k$ must vanish and, for the purpose of diagonalizing $H$, the infinite-dimensional bosonic Hilbert space splits into finite-sized blocks \cite{quantres}. The size of these blocks is determined, as mentioned above, by the number of ways to partition $M$ units into at most $N$ parts. Then, one can run elementary numerical experiments by diagonalizing the GG Hamiltonian \eqref{eq: HGG} within such individual $(N,M)$-blocks to observe, remarkably, that {\it all of its eigenvalues are nonnegative integers} \cite{quantres}. This has led us in \cite{superintegrable} to describe (\ref{eq: HGG}), somewhat metaphorically at this point, as a {\it superintegrable quantum field theory} on a circle to highlight the fact that such integer spectra are not expected from generic integrable systems. In simpler and better-controlled quantum-mechanical settings, integer spectra are well-known to be associated with superintegrable dynamics, as in the celebrated Calogero system and its relatives.

The GG Hamiltonian was shown in \cite{complexity} to also commute with the following quartic operator, whose classical counterpart may be understood straightforwardly as a member of the conserved hierarchies due to the Lax structure, see \cite{superintegrable} for further details:
\begin{equation}\label{eq: Hmin}
H_{\min}\equiv \sum^{i+j=k+l}_{i,j,k,l\ge 0} \min(i,j,k,l)\, a^\dagger_i a^\dagger_j a_k a_l + \sum_{k=0}^\infty k^2 a^\dagger_k a_k \, .
\end{equation}
A systematic exploration of the conserved hierarchies of the quantum Hamiltonian (\ref{eq: HGG}) and further associated algebraic structures was initiated in \cite{superintegrable}.
The two operators $H$ and $H_{\min}$ (which define the quartic level of these polynomial conserved hierarchies) satisfy an intertwining relation that has been instrumental in the derivation of two explicit families of eigenvectors of $H$ in \cite{superintegrable}. The approach of the present paper does not rely on this identity, but we briefly review it for additional perspective. We introduce the quantum shift operators $S$ and $S^\dagger$:
\begin{equation}
    \label{eq: relations p0 and S}
S|\eta_0,\eta_1,\cdots\rangle=|0,\eta_0,\eta_1,\cdots\rangle,\qquad S^\dagger|\eta_0,\eta_1,\eta_2,\cdots\rangle=|\eta_1,\eta_2,\cdots\rangle,
\end{equation}
where $S^\dagger$ is defined to annihilate $|\eta_0,\eta_1,\cdots\rangle$ unless $\eta_0=0$. The intertwining relation is then given by
\begin{equation}
  H_{\min} S = S(2H+H_{\min} + 2\hat M+\hat N). \label{eq:shift joint eigenvectors}
\end{equation}
Any joint eigenvector $|\Psi\rangle$ of $H$ and $H_{\min}$ in a given $(N,M)$-block thus provides an eigenvector $S |\Psi\rangle$ of $H_{\min}$ in the $(N,M+N)$-block with eigenvalue determined by \eqref{eq:shift joint eigenvectors}. As pointed out in \cite{superintegrable}, this algebraic structure can in principle be used to build the spectrum recursively. The only caveat is that the spectrum of $H_{\min}$ is highly degenerate and, for that reason, $H$ will have to be re-diagonalized within each degenerate eigenspace of $H_{\min}$ in order to proceed with the recursion.

The main result of this paper is an explicit solution for the joint integer spectrum of $H$ and $H_{\min}$. The eigenstates in a block ($N,M$) are naturally labelled by partitions $\lambda \in \mathcal{P}_N(M)$ and their eigenvalues are
\begin{equation}
    E_{\lambda} 
    = \frac12 \left(\sum_{i=1}^{d_{\lambda+1}} h_{\lambda+1}(i)^2 - \sum_{i=1}^{d_{\lambda}} h_{\lambda}(i)^2 -2M-N \right) \,,
    \label{eq: H eigenvalues intro}
\end{equation}
and
\begin{equation}
    E_{\min,\lambda} 
    = \sum_{i=1}^{d_\lambda}h_{\lambda}(i)^2 \,,
    \label{eq: Hmin eigenvalues intro}
\end{equation}
with $h_{\lambda}(i)$ and $d_\lambda$ the principal hook lengths and the Durfee size of $\lambda$, respectively, defined below in section~\ref{sec: hooks} and where $\lambda+1 \in \mathcal{P}_{N}(M+N)$ is the partition obtained by augmenting all parts of $\lambda$ by one unit. (Parts of zero size must be included into our definition of partitions, in contrast to the standard expositions of the subject, to accommodate the occupation number of mode 0. We will explain systematically the relevant definitions in section~\ref{sec: hooks}.)

The most important ingredient from \cite{superintegrable} that we will use here is the existence of a joint eigenstate of $H$ and $H_{\min}$ in every $(N,M)$-block that saturates the upper bounds 
\begin{equation}
    \frac12(\hat N-1)(\hat N+2\hat M) \geq H \quad \text{ and } \quad \hat M^2 \geq H_{\min} \, .
\end{equation}
It was shown in \cite{superintegrable} that such a vector exists uniquely in each $(N,M)$-block and must satisfy the constraints
\begin{equation}
a_{k}a_{j-k}|\psi^{\max}\rangle=a_{l}a_{j-l}|\psi^{\max}\rangle 
\end{equation}
for all $k,l,j \geq 0$. This set of constraints is solved explicitly in each block by the ``symmetric'' combination of all partitions in a given block:
\begin{equation}
    |\psi^{\max}_{N,M}\rangle=\sum_{ \lambda \in \mathcal{P}_N(M)} \left(\prod_{k=0}^\infty \frac{1}{\sqrt{\eta_k(\lambda)!}}\right)|\lambda\rangle=\sum_{ \eta \in \mathcal{M}_N(M)} \prod_{k=0}^\infty \frac{(\ad_k)^{\eta_k}}{\eta_k!}|0,0,\dots\rangle
    \label{eq: psimax}
\end{equation}
where we have traded the sum over all partitions $\lambda \in \mathcal{P}_N(M)$ for a sum over the set of all admissible occupation vectors or multiplicities $\eta = (\eta_0,\eta_1,\dots,\eta_M) \in \mathcal{M}_N(M)$ in block $(N,M)$ in the last equality. We will refrain from normalizing $|\psi^{\max}_{N,M}\rangle$, as the states \eqref{eq: psimax} satisfy the following convenient relations
\begin{equation}
a_k\,|\psi^{\max}_{N,M}\rangle =|\psi^{\max}_{N-1,M-k}\rangle , \quad a^\dagger_k\,|\psi^{\max}_{N,M}\rangle =\hat{\eta}_k|\psi^{\max}_{N+1,M+k}\rangle \,.
\label{eq: action a ad psimax}
\end{equation}

In the remainder of the paper, we  formulate and solve the diagonalization problem in the {\it Bargmann representation} where states in the Hilbert space are given by holomorphic functions of the variables $z = (z_0,z_1,\dots)$. (The solution of the closely related quantum Benjamin-Ono system has also relied on the Bargmann representation \cite{qBO}, though the strategy is not immediately related to what we are pursuing here.) This formulation is obtained from the creation–annihilation operator formalism by the following substitutions 
\begin{align}
    a_k^\dagger \rightarrow z_k,\qquad 
    a_k \rightarrow \frac{\del}{\del z_k},  \qquad
    |0,0,\dots\rangle \rightarrow 1 \, .
        \label{eq: Bargmann}
\end{align}
The operators $a$ and $\ad$ are Hermitian-conjugate with respect to the usual Bargmann inner product
\beq\label{Bprod}
\langle f|g\rangle=\frac1\pi \int  \overline{f(z)}\,g(z) \,e^{-|z|^2} d\bar zdz,
\eeq
which is straightforwardly extended to the full set of modes indexed by $k$ using the direct-product structure.
In this language, the Hamiltonians \eqref{eq: HGG} and \eqref{eq: Hmin} become second-order differential operators that we write, with $\del_k\equiv \del/\del z_k$, as
\begin{equation}
    \mathcal{H} =  \frac12\sum^{i+j=k+l}_{i,j,k,l\ge 0} z_i z_j \del_k\del_l  \, 
    \label{eq: HGG Bargmann}
\end{equation}
and 
\begin{equation}
    \mathcal{H}_{\min}  =  \sum^{i+j=k+l}_{i,j,k,l\ge 0}  \min(i,j,k,l) z_i z_j \del_k\del_l  + \sum_{k=0}^\infty k^2 z_k \del_k\,  .
    \label{eq: Hmin Bargmann}
\end{equation}
Their joint eigenstates form a hierarchy of finite-degree polynomials with,  as we shall demonstrate, integer eigenvalues.
 
The Fock states in block $(N,M)$ can be characterized in the following equivalent ways, introducing two shorthand notations for future convenience,
\begin{align}
    |\lambda\rangle = \frac{z^\lambda}{\sqrt{\lambda!}} \equiv \frac{z_0^{\eta_0}\prod_{i=1}^{N-\eta_0} z_{\lambda_i}}{\sqrt{\prod_{i=0}^\infty \eta_i!}} 
    =\frac{z^\eta}{\sqrt{\eta!}} \equiv   \prod_{i=0}^\infty\frac{z_i^{\eta_i}}{\sqrt{\eta_i!}} 
    \label{eq: Fock bargmann}
\end{align}
with $\eta \in \mathcal{M}_N(M)$ the occupation vector of the partition $\lambda \in \mathcal{P}_N(M)$. This basis generates the finite-dimensional linear space 
\beq\label{defLMN}
\mathcal{L}_{N,M}\equiv \mbox{span}\,\Big\{\prod_{k=0}^\infty z_k^{\eta_k}\,:\, \sum_k\eta_k=N,\,\sum_k k\eta_k=M\Big\}
\eeq
 of holomorphic polynomials of {\it degree} $N$ and {\it weight} $M$, such that $f(z) \in \mathcal{L}_{N,M}$ iff $\mathcal{N} f(z) = N f(z) $ and $\mathcal{M} f(z) = M f(z)$ with 
\begin{equation}
    \mathcal{N} = \sum_{k=0}^\infty z_k \del_k , \quad \text{ and } \quad  \mathcal{M}  =\sum_{k=0}^\infty k \,z_k \del_k \, ,
    \label{eq: NM bargmann}
\end{equation}
the Bargmann representation of the operators \eqref{eq: NMq}.

The highest energy state \eqref{eq: psimax} in each $(N,M)$-block takes the form 
\begin{equation}
    \psmax{N}{M} \,\,= \sum_{\lambda \in \mathcal{P}_N(M)} \frac{z^{\lambda}}{\lambda!} \,\,\,= \sum_{\eta \in \mathcal{M}_N(M)} \prod_{i=0}^\infty\frac{z_i^{\eta_i}}{\eta_i!} \,,
    \label{eq: psimax bargmann}
\end{equation}
with eigenvalues
\begin{equation}
    e_{\scriptscriptstyle{N,M}} = \frac12( N-1)( N+2 M) \quad \text{ and } \quad  e_{\scriptscriptstyle{N,M}}^{\min} = M^2 \, ,
    \label{eq: energies highest state}
\end{equation}
for $\mathcal H$ and $\mathcal H_{\min}$ respectively.
These states arise as the expansion coefficients (with respect to $c$ and $p$) of a following simple coherent state 
\begin{equation}
    \Psi_{c,p}(z) \equiv \exp\left[c \sum_{k=0}^\infty p^k z_k\right] = 1+ \sum_{N > 0}^\infty \sum_{M \geq 0}^\infty \psmax{N}{M}  c^N p^M  \, ,
   \label{eq: coherent psi max}
\end{equation}
associated with the single-soliton configuration $\al_k=cp^k$ or
\begin{equation}
    u(\ze) = \frac{c}{1-p \ze} \, 
    \label{eq: lowest invariant manifold}
\end{equation}
in the classical theory (\ref{cuSz}).
The coherent states \eqref{eq: coherent psi max} define exact joint eigenstates for the operators $H-\frac12(\hat N-1)(\hat N+2\hat M)$ and $H_{\min}-\hat M^2$ with eigenvalues equal to zero. The same is true of the single-soliton configurations (\ref{eq: lowest invariant manifold}) in the classical theory (\ref{cuSz}). Further connections between families of eigenstates of $H$ and $H_{\min}$ and higher-rank classical invariant manifolds (\ref{classinv})
are expected, some of which will be illuminated from the eigenbasis construction to be presented here.

The highest energy states will provide the foundation for a complete eigensolution, somewhat in the spirit of supersymmetric quantum mechanics and the factorization method. The intuition behind the construction we shall present can be loosely explained as follows: since the Bargmann-representation operators $\mathcal{H}$ and $\mathcal{H}_{\min}$ are second-order differential operators, they do not satisfy the Leibniz rule  in the sense that $\mathcal{H}(fg)\ne \mathcal{H}(f)g+f\mathcal{H}(g)$. We shall show, however, that a certain ordering on polynomials can be introduced such that an appropriate variation of the Leibniz rule is in fact satisfied by both $\mathcal{H}$ and $\mathcal{H}_{\min}$, up to lower-order contributions.
This formulation will naturally allow for using products of eigenvectors to construct new eigenvectors, again, up to lower-order contributions, and the top eigenvectors (\ref{eq: psimax bargmann}) will emerge as the main ingredient for such products. With all these accessories at hand, we will be able to define an explicit basis whose vectors are joint eigenvectors of $\mathcal{H}$ and $\mathcal{H}_{\min}$, up to lower-order contributions. But this will automatically imply that $\mathcal{H}$ and $\mathcal{H}_{\min}$ are upper-triangular in this basis, when ordered according to our definition, from which it follows that the true eigenvectors can be recovered by a straightforward Gram-Schmidt process (with no full-blown diagonalization required), while the eigenvalues can be read off immediately from the diagonal entries of this upper-triangular matrix representation of $\mathcal{H}$ and $\mathcal{H}_{\min}$, and they are manifestly integers.

The remainder of the paper is structured as follows: In the next section, we give an explicit basis for $\mathcal{L}_{N,M}$, constructed from specific products of highest energy states \eqref{eq: psimax bargmann} that are in one-to-one correspondence with partitions of $\mathcal{P}_N(M)$. This construction, in combination with a reverse lexicographic ordering on the indices of these products, provides a natural grading on $\mathcal{L}_{N,M}$ in terms of which $\mathcal{H}$ and $\mathcal{H}_{\min}$ display a block-upper-triangular structure, derived in section~\ref{sec: staircase}. Combining these two ingredients will then allow us to characterize the eigenvectors and derive an explicit formula for the integer spectrum of both operators in section~\ref{sec: eigenbasis}. We will conclude with a discussion.


\section{An ordered basis for $(N,M)$-blocks}
\label{sec: hooks}

We start by introducing a totally ordered set of states in $\mathcal{L}_{N,M}$ defined by (\ref{defLMN}). This ordering will be formulated in relation to polynomials constructed as products of the highest energy states \eqref{eq: psimax bargmann} in lower blocks. As we shall discuss, these states naturally appear as expansion coefficients of polynomials in $c$, $c_r$ and $p_r$ for the quantum coherent states corresponding to the classical multisoliton configurations \eqref{classinv}. They furthermore form an overcomplete basis for $\mathcal{L}_{N,M}$. A reverse lexicographic ordering on these states will naturally suggest a particular basis for $\mathcal{L}_{N,M}$ in section \ref{subsec: basis}, whose elements are in one-to-one correspondence with partitions in $\mathcal{P}_N(M)$. In section \ref{sec: staircase}, we shall see that, with this ordering of the basis states, the Hamiltonians will respect a very useful approximate Leibniz rule up to lower-order terms.

\subsection{A totally ordered set of states}

We fix the values $(N,M)$ and consider the set $X_{N,M}$ of all decompositions $N = \sum_{i=1}^d n_i$ and $M = \sum_{i=1}^d m_i$ for any $d>0$, with $n_i>0$ and $m_i\geq 0$, obeying the following sorting rule: either $m_{i} \geq m_{i+1} $, or if $m_i = m_{i+1}$, one must have $n_i \geq n_{i+1}$. We store this data as lists
\begin{equation}
    x = \{m_1,n_1,\cdots,m_d,n_d \} \in X_{N,M} \, .
    \label{eq: list for ordering}
\end{equation}
We then define the following set of states 
 \begin{equation}
     \phi_{x \in X_{N,M}} = \phi_{\{n_i,m_i\}}  \equiv \prod_{i=1}^{d} \psmax{n_i}{m_i}
     \label{eq: basis products psimax}
 \end{equation} 
constructed as a string of highest energy states \eqref{eq: psimax bargmann} in the blocks with labels ($n_i,m_i$), with nonincreasing momentum $m_i$. Note that such product-strings of highest energy states \eqref{eq: basis products psimax} naturally emerge as the coefficients of symmetric polynomials in the classical variables $c$, $c_r$, $p_r$ from expanding the standard coherent states
\begin{equation}
    \Psi_R(c, \{c_r\},\{p_r\}) \equiv \exp\left[c z_0 + \sum_{r=1}^R c_r \sum_{k=0}^\infty p_r^k z_k\right]   \, ,
   \label{eq: coherent multisoliton}
\end{equation}
built from the classical multisoliton configurations \eqref{classinv} with $R \rightarrow \infty$, as seen by isolating the terms with total particle number $N$ and total momentum $M$.
 
To sort the states \eqref{eq: basis products psimax}, we note that the set $X_{N,M}$ admits a total order that follows from a reverse lexicographic order on the data ${\{n_i,m_i\}}$. 
Using the list representation \eqref{eq: list for ordering}, we define the ordering on the states \eqref{eq: basis products psimax} as the reverse lexicographic order on $x$:
\begin{equation}
    \phi_x \prec \phi_{x'}  \quad \leftrightarrow \quad x >_{\text{lex}} x' \, .
    \label{eq: ordering phis}
\end{equation}
Specifically, we first order states according to decreasing values of $m_1$. If those are equal, we compare the second element of the list $x$ and sort according to decreasing values of $n_1$. If the first pair $(m_1,n_1)$ of two states \eqref{eq: basis products psimax} is identical, we compare $m_2$ and sort in decreasing order, etc. Note that we never compare a pair ($m_j,n_j$) to an empty pair, since all $n_i$ and $m_i$ in $x$ sum up to $N$ and $M$, and one only needs to compare a pair at some value $j$ if all preceding pairs are equal. The highest energy state $\psmax{N}{M}$ is evidently the lowest state for the ordering \eqref{eq: ordering phis} in block ($N,M$).

When all $n_i = 1$, the products \eqref{eq: basis products psimax} are simply the unnormalized Fock states \eqref{eq: Fock bargmann}. It is therefore clear that the set of states \eqref{eq: basis products psimax} is overcomplete for the subspace $\mathcal{L}_{N,M}$. In particular, consider the nested subspaces 
\begin{equation}
    \Phi^{N,M}_x \equiv \operatorname{span}\{ \phi_{x'} : x' \in X_{N,M} \text{ and }    \phi_{x'} \preceq \phi_x \}
    \label{eq: span of strings}
\end{equation}
with, by construction, $\Phi^{N,M}_x \subseteq \Phi^{N,M}_{x'}$ if $x \geq_{\text{lex}} x'$.
If $ x_{\text{top}}$ is the sequence \eqref{eq: list for ordering} of highest order for a given block ($N,M$), such that the span \eqref{eq: span of strings} is considered over all states \eqref{eq: basis products psimax}, one obtains the linear space $\mathcal{L}_{N,M} = \Phi^{N,M}_{x_{\text{top}}}$.

The aim of the remainder of this section is to demonstrate that the reverse lexicographic order on $x$ selects a particular subset of states \eqref{eq: basis products psimax} that, as we discuss in section~\ref{sec: staircase}, forms a useful basis in $\mathcal{L}_{N,M}$ for the diagonalization of $\mathcal{H}$ and $\mathcal{H}_{\min}$. The idea will be to identify the states $\phi_{x}$ that define new independent directions in $\mathcal{L}_{N,M}$ in the nested construction of the subspaces $\Phi^{N,M}_x$. We are thus looking for a basis that is as low as possible in our ordering (to complement the trivial Fock basis that sits higher with respect to our ordering). The underlying structure that controls the increase in the dimension of the subspaces \eqref{eq: span of strings} by the (ordered) addition of a state $\phi_{x}$ can be traced back to properties of partitions and their hook reductions, which we will now review.

\subsection{Young diagrams and hook reduction}

We proceed to introduce the terminology of Young diagrams, hooks, principal hook lengths, appearing in the eigenspectra \eqref{eq: H eigenvalues intro} and \eqref{eq: Hmin eigenvalues intro}, and the hook reduction of a partition. These are elementary concepts from the theory of integer partitions and we refer to standard textbooks on the subject  \cite{partitions1,partitions2} for more details.

\begin{figure}[t]
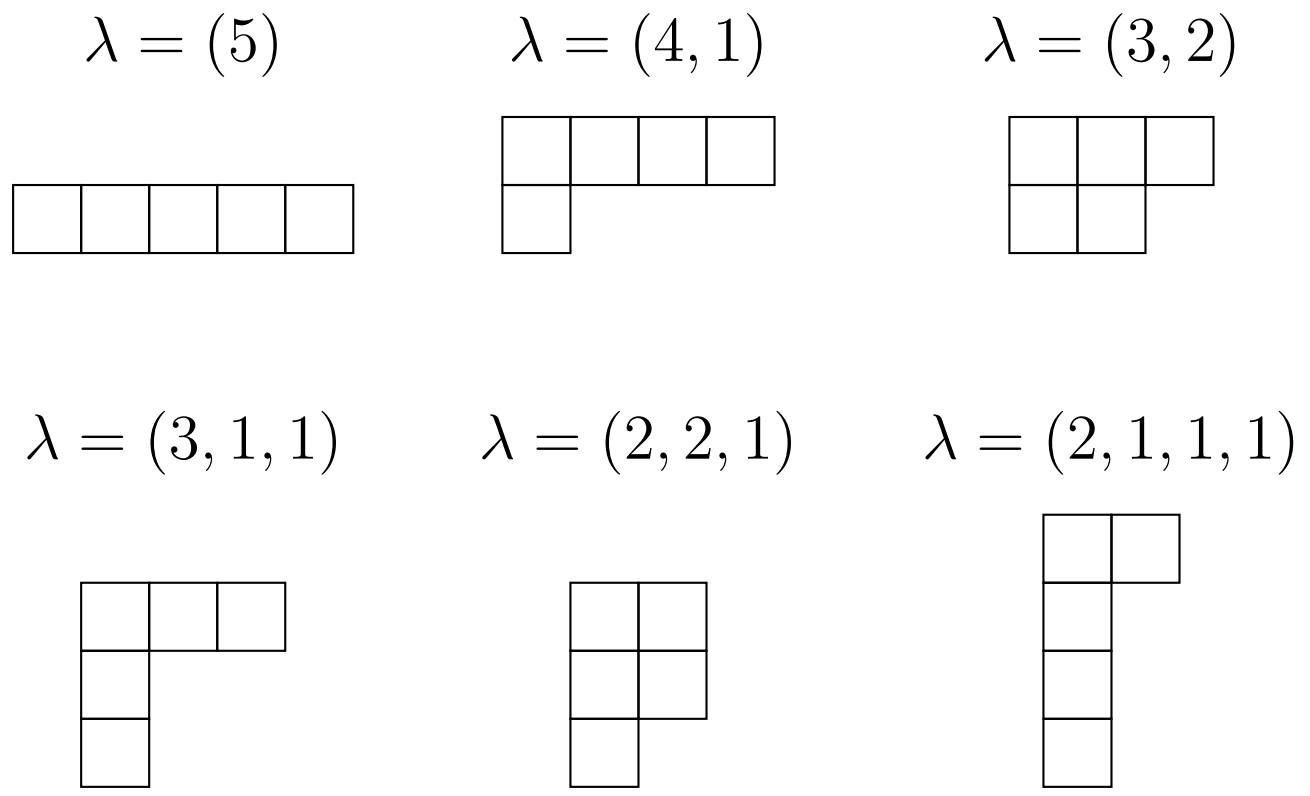

\centering
\begin{center}
\begin{tabular}{c@{\qquad}c@{\qquad}c}
$\lambda=(5)$ &
$\lambda=(4,1)$ &
$\lambda=(3,2)$ \\[6pt]
\youngdiagram{5} &
\youngdiagram{4,1} &
\youngdiagram{3,2}
\\[24pt]
$\lambda=(3,1,1)$ &
$\lambda=(2,2,1)$ &
$\lambda=(2,1,1,1)$ \\[6pt]
\youngdiagram{3,1,1} &
\youngdiagram{2,2,1} &
\youngdiagram{2,1,1,1}
\end{tabular}
\end{center}
\caption{Young diagrams for the partitions of $M=5$ with at most $N=4$ parts.}
\label{tab: Young diagrams}
\end{figure}

Consider the partitions $\lambda = (\lambda_1 , \lambda_2 , \dots , \lambda_\ell)$ of $M$ into at most $N$ parts with $\lambda_1 \geq \lambda_2 \geq \dots \geq \lambda_\ell \geq 1$, i.e. $\mathcal{P}_{N}(M) =\{\lambda \vdash M : \ell(\lambda)\leq N\}$, with $\ell(\lambda)$ the number of nonzero parts. The notation $\lambda \vdash M $ means that $\sum_{i=1}^\ell \lambda_i=M$. A Young diagram is a visual representation of a partition, with a total of $M$ boxes arranged in rows aligned along the left side. The number of boxes in the $i^\text{th}$ row is given by $\lambda_i$. An illustration of all the Young diagrams for the block $(N,M)=(4,5)$ is provided in Fig.~\ref{tab: Young diagrams}.

It is important to emphasize that {\it parts of zero size} are not considered in the standard textbook theory of partitions and
 are not explicitly shown in the Young diagram. But they must be incorporated in our analysis to correctly capture $\eta_0$, the occupation number of mode 0.
The latter can be deduced from $\eta_0(\lambda) = N-\ell(\lambda)$,  chosen precisely to keep the total occupation number $\sum_{k=0}^\infty\eta_k$ equal $N$.
When we say `partition' below, we mean standard partitions `augmented' by $\eta_0$ parts of size 0 (which could in principle be depicted as $\eta_0$ additional dots below  the first column in the Young diagram). To represent the same, we could equivalently increase each part by $+1$, including the parts of zero size, obtaining a conventional partition (without parts of zero size) of $M+N$ into exactly $N$ parts that we have already briefly mentioned above, denoted symbolically ``$\lambda+1$'' in the discussion below (\ref{eq: Hmin eigenvalues intro}).

\begin{figure}[b]
\centering

\begin{tabular}{ccc}


\begin{tikzpicture}[scale=0.4]

\foreach \x in {1,...,6} {
    \fill[hookone] (\x-1,0) rectangle (\x,1);
}
\foreach \y in {1,...,5} {
    \fill[hookone] (0,-\y) rectangle (1,-\y+1);
}

\foreach \x in {2,...,4} {
    \fill[hooktwo] (\x-1,-1) rectangle (\x,0);
}
\foreach \y in {2,...,5} {
    \fill[hooktwo] (1,-\y) rectangle (2,-\y+1);
}

\foreach \x in {3,...,4} {
    \fill[hookthree] (\x-1,-2) rectangle (\x,-1);
}
\foreach \y in {3,...,4} {
    \fill[hookthree] (2,-\y) rectangle (3,-\y+1);
}

\fill[hookfour] (3,-3) rectangle (4,-2);

\foreach \y/\length in {
    0/6,
    1/4,
    2/4,
    3/4,
    4/3,
    5/2
} {
    \foreach \x in {1,...,\length} {
        \draw (\x-1,-\y) rectangle (\x,1-\y);
    }
}

\end{tikzpicture}
&
\begin{tikzpicture}[scale=0.4]

\foreach \x in {1,...,5} {
    \fill[hookone] (\x-1,0) rectangle (\x,1);
}
\foreach \y in {1,...,6} {
    \fill[hookone] (0,-\y) rectangle (1,-\y+1);
}

\foreach \x in {2,...,5} {
    \fill[hooktwo] (\x-1,-1) rectangle (\x,0);
}
\foreach \y in {2,...,4} {
    \fill[hooktwo] (1,-\y) rectangle (2,-\y+1);
}

\foreach \x in {3,...,4} {
    \fill[hookthree] (\x-1,-2) rectangle (\x,-1);
}
\foreach \y in {3,...,4} {
    \fill[hookthree] (2,-\y) rectangle (3,-\y+1);
}

\fill[hookfour] (3,-3) rectangle (4,-2);

\foreach \y/\length in {0/5,1/5,2/4,3/4,4/3,5/1,6/1} {
    \foreach \x in {1,...,\length} {
        \draw (\x-1,-\y) rectangle (\x,1-\y);
    }
}

\end{tikzpicture}
&
\begin{tikzpicture}[scale=0.4]

\foreach \x in {1,...,8} {
    \fill[hookone] (\x-1,0) rectangle (\x,1);
}
\foreach \y in {1,...,3} {
    \fill[hookone] (0,-\y) rectangle (1,-\y+1);
}

\foreach \x in {2,...,6} {
    \fill[hooktwo] (\x-1,-1) rectangle (\x,0);
}
\foreach \y in {2,...,3} {
    \fill[hooktwo] (1,-\y) rectangle (2,-\y+1);
}

\foreach \x in {3,...,5} {
    \fill[hookthree] (\x-1,-2) rectangle (\x,-1);
}
\fill[hookthree] (2,-3) rectangle (3,-2);

\fill[hookfour] (3,-3) rectangle (4,-2);

\foreach \y/\length in {0/8,1/6,2/5,3/4} {
    \foreach \x in {1,...,\length} {
        \draw (\x-1,-\y) rectangle (\x,1-\y);
    }
}
\end{tikzpicture}
\end{tabular}
\caption{The Young diagrams for the partitions ($6,4,4,4,3,2$), ($5,5,4,4,3,1,1$) and $(8,6,5,4)$ with their four principal hooks highlighted. These three diagrams have identical principal hook lengths ($11,7,4,1$) and represent valid partitions of $M = 23$ as long as $N\geq6$, $N\geq7$ and $N\geq4$, respectively.}
\label{tab: principal hooks}
\end{figure}
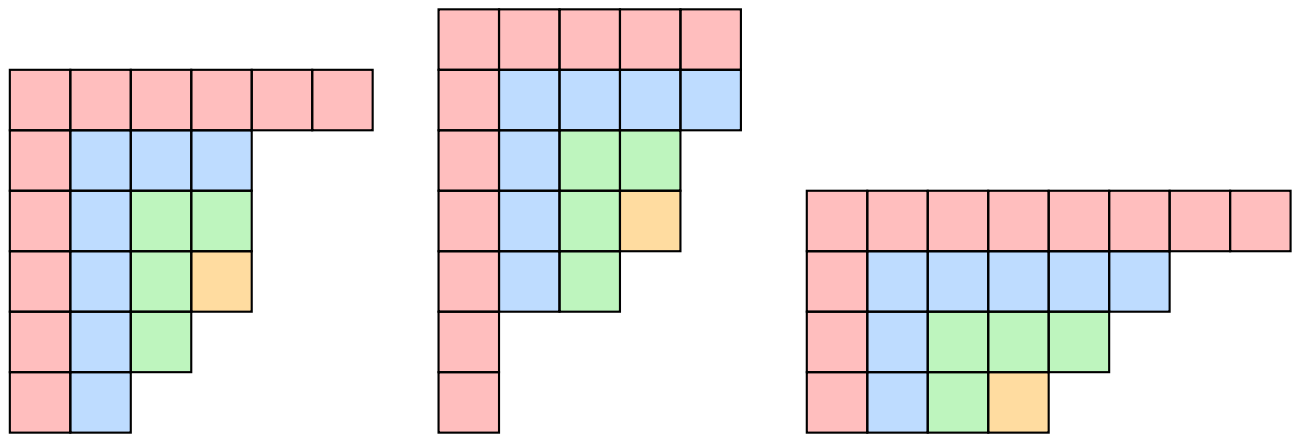

A box inside a Young diagram is specified by its row and column number $(r,c)$. The \textit{hook} of a box at ($r,c$) consists of the inverted $L$-shaped collection of boxes that contains the box itself, as well as all the boxes to the right of and below $(r,c)$. The hook length at $(r,c)$ is the number of boxes constituting the corresponding hook. The principal hooks, defined as the hooks along the diagonals where $r=c=i$, partition the entire Young diagram. Their lengths are known as \textit{principal hook lengths} $h_\lambda(i)$. We shall refer to the maximal diagonal index $i$ of a Young diagram as $d_{\lambda}$, the Durfee size, which corresponds to the size of the largest square that fits into a Young diagram. Note that the principal hook lengths do not determine the partition $\lambda$, since a box in the $i^{\text{th}}$ hook can be moved from the respective row to the column (or vice versa) without changing the length, as long as $ \lambda_i > \lambda_{i+1}$ and $\lambda'_i < \lambda'_{i-1}$ with $\lambda'$ the conjugate partition (or $ \lambda_i < \lambda_{i-1}$  and $\lambda'_i > \lambda'_{i+1}$). Three partitions with same principal hook lengths are shown in Fig.~\ref{tab: principal hooks}. In order to fully specify a partition, it is convenient to introduce Frobenius coordinates $\lambda = (a_1,\dots, a_{d_\lambda}| b_1, \dots, b_{d_\lambda})$, with $a_i$ and $b_i$ respectively the number of boxes to the right and to the bottom of the diagonal $(i,i)$, such that $h_\lambda(i) = a_i + b_i +1$.

For a partition $\lambda \in \mathcal{P}_{N}(M)$ of {\it length} $\ell(\lambda) \leq N$, given by the number of rows in the Young diagram, the number of nonzero parts is, by definition, $\ell(\lambda)$ and the zero-part multiplicity is $\eta_0(\lambda) = N- \ell(\lambda)$. The {\it hook reduced partition} $\mu$ is obtained by removing the outer principal hook of $\lambda$, i.e. $r(\lambda) = \mu \equiv (\lambda_2-1,\lambda_3-1,\cdots,\lambda_{\ell(\lambda)}-1)$. The zero parts of $\lambda$ are discarded in this procedure. The zero-part multiplicity of $\mu$ is given by $\eta_0(\mu) = \eta_1(\lambda)$ (except if $\eta_1(\lambda) = M$, in which case $\eta_0(\mu) = \eta_1(\lambda) -1 $).  The number of boxes of the Young diagram of $\mu$ is  $M_1 =  \sum_i \mu_i = M-h_{\lambda}(1) < M$ and the number of parts is given by $N_1 = \ell(\lambda) -1 = b_1(\lambda) < N$. The reduced partition $\mu$ belongs to $\mathcal{P}_{N_1}(M_1)$. The number of nonzero parts of $\mu$ thus relates to the original length as $\ell(\mu) \leq \ell(\lambda)-1$.

The process of hook-reducing a partition can be applied iteratively. One starts with a partition $\lambda \in \mathcal{P}_{N}(M)$, with corresponding Young diagram and zero-part multiplicity $\eta_0(\lambda) = N - \ell(\lambda)$. Each hook-reduction step $i$ defines $\lambda^{(i)} = r(\lambda^{(i-1)})$ (with the convention $\lambda^{(0)} \equiv \lambda$), which is visually represented by stripping off the outer principal hook of the Young diagram of partition $\lambda^{(i-1)} \in \mathcal{P}_{N_{i-1}}(M_{i-1})$, with the convention  $(N_0,M_0) = (N,M)$, and defining $N_i$ by the assignment of a multiplicity $\eta_0(\lambda^{(i)}) = \ell(\lambda^{(i-1)})-\ell(\lambda^{(i)})-1$ to its zero parts. This recursive process terminates when reaching the final partition with no remaining parts, which is $\lambda^{(d_{\lambda})}$ if $N_{d_{\lambda}} = 0$ or $\lambda^{(d_{\lambda}+1)}$ otherwise, with $d_\lambda$ the Durfee size of $\lambda$.

\subsection{The hook-reduced basis for $\mathcal{L}_{N,M}$}
\label{subsec: basis}

The hook-reduction process can be used to naturally assign one state of the form \eqref{eq: basis products psimax} to each partition $\lambda \in \mathcal{P}_N(M)$. To see this, we consider the blocks $(N_i,M_i)$ in which the hook-reduced partitions of $\lambda$, i.e.~$\lambda^{(i)} = r^i(\lambda)$ with $ r^0(\lambda) \equiv \lambda $, are defined and note that the principal hook lengths $h_{\lambda}(i) = M_{i-1}-M_i$ represent a decreasing sequence that partitions $M$ that can thus be identified with the sequence $m_i$ in \eqref{eq: list for ordering} and, likewise, the differences $n_i \equiv N_{i-1}-N_{i}$ partition $N$. One can thus associate the following product of highest energy states to $\lambda$:
 \begin{equation}
     \phi_{\lambda} \equiv \prod_{i=1}^{d_\lambda+1} \psmax{N_{i-1}-N_i}{M_{i-1}-M_i}
     \label{eq: basis products psimax final}
 \end{equation} 
 with the conventions $\psmax{0}{0}=1$ and $N_{d_\lambda+1} = M_{d_\lambda+1} =0$. We shall refer below to the collection of states \eqref{eq: basis products psimax final} as 
 \begin{equation}
     B_{N,M} \equiv \{ \phi_\lambda : \lambda \in \mathcal{P}_N(M) \} \, .
     \label{eq: basis for ordering}
 \end{equation}
 
 For future reference, we investigate how to characterize a sequence $\{n_i,m_i\}$ that emerges from such a hook-reduction process. We start by interpreting the values $m_i$ as principal hook lengths, and introduce their corresponding Frobenius coordinates, so that $m_i = a_i + b_i+1$. It is also straightforward to see from a Young diagram representation that variables $n_i$ for $i>1$ are related to the leg lengths $b_i$ as $n_i =  b_{i-1} - b_i$. (At $i=1$, one has the relation $n_1 = \eta_0(\lambda) +1 >0 $ but this will not be needed.) There are only two conditions for a sequence of Frobenius coordinates to be consistent:
 \begin{itemize}
     \item the inner leg should fit into the outer leg, such that $b_{i} > b_{i+1} \leftrightarrow n_{i+1} >0$
     \item the inner arm should fit into the outer arm, such that 
     \begin{equation}
         a_{i} > a_{i+1} \leftrightarrow  m_i> n_{i+1}+ m_{i+1}
         \label{eq: condition hook reduced}
     \end{equation}
 \end{itemize}
 for $i\geq 1$. 
From these conditions, it is clear that the correspondence \eqref{eq: basis products psimax final} between partitions for the block $(N,M)$ and the states \eqref{eq: basis products psimax} is not one-to-one, as most assignments $\{n_i,m_i\}$ in \eqref{eq: basis products psimax} do not originate from the hook-reduction of a partition of the block ($N,M$). The remainder of this section demonstrates that the states \eqref{eq: basis products psimax final} are precisely the elements that increase the dimension of $ \Phi^{N,M}_x$ \eqref{eq: span of strings}. They thus provide a natural basis within the overcomplete set of states \eqref{eq: basis products psimax} with respect to the reverse lexicographic order on \eqref{eq: list for ordering}.  
 
The key identity that distinguishes the states \eqref{eq: basis products psimax final} from the other products can be formulated as the following statement about products of two highest energy states: When $n_2+m_2 \geq m_1$, one finds that the product $\psmax{n_1}{m_1}\psmax{n_2}{m_2}$
is supported on two-factor states \eqref{eq: basis products psimax} of lower order within the block $( n_1 + n_2,m_1+m_2)$:
\begin{align}
    \psmax{n_1}{m_1}\psmax{n_2}{m_2} \in \operatorname{span}\{ \psmax{n_1'}{m_1'}&\psmax{n_2'}{m_2'} : n_1+n_2 = n_1’+n_2’\,, \nonumber  \\  &\quad  m_1+m_2 = m_1’+m_2’ \text{ and }  (m_1’,n_1') >_{\text{lex}}(m_1,n_1) \} \, .
    \label{eq: support two-soliton}
\end{align} Note that the inequality in \eqref{eq: support two-soliton} is interpreted in terms of the reverse lexicographic order (i.e.~either $m_1'>m_1$ holds or $m_1'=m_1$ and $n_1'>n_1$). This statement, proven in appendix~\ref{app: support}, implies that states \eqref{eq: basis products psimax} for which any one pair of subsequent highest energy state factors satisfies 
\begin{equation}
    n_{i+1}+m_{i+1} \geq m_{i}
    \label{eq: condition non hook reduced}
\end{equation}
can be expanded in lower order states \eqref{eq: basis products psimax} (possibly involving a reordering of highest energy factors in certain terms, which can only lower the ordering) and therefore do not determine new independent directions when included in the nested subspaces \eqref{eq: span of strings}.

The key observation is that pairs of highest energy states originating from the process of hook-reduction of a partition never satisfy \eqref{eq: condition non hook reduced}, due to \eqref{eq: condition hook reduced}. Indeed, if \eqref{eq: condition non hook reduced} applies to any one pair of highest energy states, then this implies that indices of the two considered highest energy states cannot be associated to a pair of subsequent principal hooks in an allowed Young diagram, with the principal hook encoded by $\psmax{n_{i+1}}{m_{i+1}}$ representing the inner and $\psmax{n_i}{m_i}$ the outer hook. Moreover, states \eqref{eq: basis products psimax} that are not of the form \eqref{eq: basis products psimax final} must have at least one pair with \eqref{eq: condition non hook reduced}, since otherwise the corresponding sequence \eqref{eq: list for ordering} would be associated with a partition by \eqref{eq: condition hook reduced} and the state would be of the form \eqref{eq: basis products psimax final}.

We thus find that states \eqref{eq: basis products psimax} \textit{not} corresponding to hook-reduced partitions as in \eqref{eq: basis products psimax final} are redundant in the span \eqref{eq: span of strings} when the states are ordered with the reverse lexicographic order on \eqref{eq: list for ordering}. We have established earlier that the complete set \eqref{eq: basis products psimax} spans $\mathcal{L}_{N,M}$. In addition, the states \eqref{eq: basis products psimax final} are in one-to-one correspondence with partitions in $\mathcal{P}_{N}(M)$ so that $|B_{N,M}| = \dim \mathcal{L}_{N,M}$. These two facts combined then directly imply linear independence of the subset \eqref{eq: basis for ordering} and demonstrate that $B_{N,M}$ defines a basis for $\mathcal{L}_{N,M}$.

\section{The triangular structure of $\mathcal{H}$ and $\mathcal{H}_{\min}$} 
\label{sec: staircase}

Our next aim is to show that the Hamiltonians $\mathcal H$ and $\mathcal{H}_{\min}$ are upper-triangular in the basis $B_{N,M}$ with ordering \eqref{eq: ordering phis}. Introducing the following graded subspaces on $\mathcal{L}_{N,M}$
\begin{equation}
      \Phi^{N,M}_{\prec \lambda} \equiv \operatorname{span}\{ \phi_{\lambda'} \in B_{N,M} :\phi_{\lambda'} \prec \phi_{\lambda}  \}
    \label{eq: grading}
\end{equation}
with $\lambda \in \mathcal{P}_N(M)$, we aim to show that the Hamiltonians preserve the grading:
\begin{equation}
    \mathcal{H} : \Phi^{N,M}_{\prec \lambda} \rightarrow \Phi^{N,M}_{\prec \lambda} \quad \text{and} \quad  \mathcal{H}_{\min} : \Phi^{N,M}_{\prec \lambda} \rightarrow \Phi^{N,M}_{\prec \lambda}  \, .
    \label{eq: H Hmin preserve filtration}
\end{equation}
This will be achieved by applying the product rule to the Bargmann representations \eqref{eq: HGG Bargmann} and \eqref{eq: Hmin Bargmann} expressed in terms of second order differential operators acting on a basis element $\phi_\lambda$, factored as
\begin{equation}
    \phi_\lambda = \psmax{n_1}{m_1}
 \phi_{r(\lambda)} \,,
 \label{eq: recursive factorization phi lambda}
 \end{equation}
 with $r(\lambda)$ the hook-reduction of $\lambda$ and $\phi_{r(\lambda)} \in \mathcal L_{N_1,M_1}$ an element of $B_{N_1,M_1}$, see \eqref{eq: basis products psimax final} for a reminder of the definitions of the different types of indices.
We will show that the cross-terms (single-derivative contributions) that usually prevent second-order differential operators from satisfying the Leibniz rule have support on strictly lower-order elements in $B_{N,M}$ compared to $\phi_\lambda$. We shall then use the resulting approximate Leibniz rule (up to lower orders) recursively to $\phi_{r^i(\lambda)}$, for $i>0$, to demonstrate the upper triangularity of the two Hamiltonians with respect to the ordered basis \eqref{eq: basis for ordering}.

\subsection{Approximate Leibniz rule for $\mathcal H$}
As a first step, we will show that
\begin{align}    \label{eq: selfsimilarity hgg}
    &\mathcal{H}(\phi_\lambda) 
    =  \mathcal{H}(\psmax{n_1}{m_1})  \phi_{r(\lambda)}  + \psmax{n_1}{m_1} \mathcal{H}( \phi_{r(\lambda)})   \,
\end{align}
holds up to terms with support in $\Phi^{N,M}_{\prec \lambda}$.
It will be convenient to prove this approximate Leibniz rule by interpreting \eqref{eq: recursive factorization phi lambda} as the product of $\phi_{r(\lambda)}$ and the single-soliton coherent state \eqref{eq: coherent psi max},  projected to the block ($N,M$). For future convenience, we list the following useful identities:
\beq
     \del_k \Psi_{c,p}(z) = cp^k  \Psi_{c,p}(z),   \qquad 
     z_k =  \psmax{1}{k} ,
      \qquad 
    \del_k \psmax{N}{M} =  \psmax{N-1}{M-k},
     \label{eq: a on psi max bargmann} 
\eeq
the first and third identities being equivalent to \eqref{eq: action a ad psimax}.
We observe that the Bargmann formulation of $a^\dagger$ can be interpreted as a highest energy state in a block with $N=1$.  
We then write 
\begin{align}
    \mathcal{H}(\Psi_{c,p}(z)\phi_{r(\lambda)}) &= \frac12\sum^{i+j=k+l}_{i,j,k,l\ge 0} z_i z_j \del_k\del_l\left( \Psi_{c,p}(z) \phi_{r(\lambda)} \right) \nonumber \\
    &= \mathcal{H}(\Psi_{c,p}(z))\phi_{r(\lambda)}  + \Psi_{c,p}(z) \mathcal{H}(\phi_{r(\lambda)}) + \Psi_{c,p}(z) \sum^{i+j=k+l}_{i,j,k,l\ge 0}  c p^l  z_i z_j  \del_k \phi_{r(\lambda)} \, .
    \label{eq: leibniz coherent}
\end{align}

The goal is to demonstrate that the third term in \eqref{eq: leibniz coherent} involves contributions to $\mathcal{L}_{N,M}$ of strictly lower order \eqref{eq: ordering phis} compared to $\phi_\lambda$. (One should project this expression on $\mathcal{L}_{N,M}$ to make this statement precise.) This will follow immediately if the factor multiplying the coherent state in that term can be written as a linear combination of states \eqref{eq: basis products psimax} in the blocks $\mathcal{L}_{N',M'}$, with parameters of strictly lower lexicographic order $(M',N') <_{\text{lex}} (M_1,N_1)$. (Remember $\phi_{r(\lambda)} \in \mathcal{L}_{N_1,M_1}$.)
To verify this, we write
\begin{align}
\sum^{i+j=k+l}_{i,j,k,l\ge 0} p^l z_i z_j \del_k\phi_{r(\lambda)}   &= \sum^{i+j\geq k}_{i,j,k\ge 0} p^{i+j-k} z_i z_j \del_k \phi_{r(\lambda)} \nonumber \\&= \sum_{i,j,k\ge 0} p^{i+j-k} z_i z_j \del_k \phi_{r(\lambda)} - \sum^{i+j< k}_{i,j,k\ge 0} p^{i+j-k}  z_i z_j \del_k \phi_{r(\lambda)} \, .
\label{eq: HGG c1}
\end{align}
The sums in the first term of \eqref{eq: HGG c1} are now unconstrained, and one can use the identity 
\begin{equation}
c^2\Psi_{c,p}(z) \sum_{i,j \ge 0} p^{i+j} z_i z_j  =  \mathcal N( \mathcal N-1) \Psi_{c,p}(z) \,,
    \label{eq: identity N^2}
\end{equation}
with $\mathcal{N}$ defined by (\ref{eq: NM bargmann}),
to remove the factors of $z$. This follows immediately from the definition \eqref{eq: NM bargmann}, with the linear piece taking care of normal ordering the factors of $z$ and the derivatives:
\begin{align}
     \mathcal N(\mathcal N-1) \Psi_{c,p}(z) &= \sum_{i=0}^\infty z_i \del_i \left[ \left( \sum_{j=0}^\infty z_j \del_j -1 \right)\Psi_{c,p}(z) \right]  = \sum_{i,j=0}^\infty z_i z_j \del_i \del_j \Psi_{c,p}(z) \nonumber \\&=c^2\Psi_{c,p}(z) \sum_{i,j \ge 0} p^{i+j} z_i z_j. \nonumber
\end{align} The action of a derivative $\del_k$ on $\phi_{r(\lambda)}$ can be worked out using \eqref{eq: a on psi max bargmann} and leads to a linear combination of products of highest energy states with a total degree $N'=N_1-1$ and weight $M'=M_1-k$, confirming  $(M',N') <_{\text{lex}} (M_1,N_1)$. When multiplied with \eqref{eq: identity N^2}, with $\mathcal N$ acting diagonally in blocks of fixed value $N$, we thus find that the first term in \eqref{eq: HGG c1} yields contributions to the block $(N,M)$ of strictly lower reverse lexicographic order \eqref{eq: ordering phis} compared to $\phi_\lambda$.

By a similar argument that uses the second and third identity of \eqref{eq: a on psi max bargmann},  the second type of contributions to \eqref{eq: HGG c1} can be written as linear combinations of products \eqref{eq: basis products psimax} in $\mathcal{L}_{N',M'}$, with $M' < M_1$,  due to the constraint on the indices in the sums $k-i-j >  0$. The entire third contribution to \eqref{eq: leibniz coherent}, projected into the block $(N,M)$, is thus of strictly lower order than $\phi_\lambda$. This confirms that the approximate Leibniz rule \eqref{eq: selfsimilarity hgg} holds up to terms belonging to $\Phi^{N,M}_{\prec \lambda}$.

\subsection{Approximate Leibniz rule for $\mathcal H_{\min}$}

We now proceed to verify a similar statement for $\mathcal{H}_{\min}$:
\begin{align}    \label{eq: selfsimilarity hmin}
    &\mathcal{H}_{\min}(\phi_\lambda) 
    =  \mathcal{H}_{\min}(\psmax{n_1}{m_1})  \phi_{r(\lambda)}  + \psmax{n_1}{m_1} \mathcal{H}_{\min}( \phi_{r(\lambda)})   \,
\end{align}
up to terms with support in $\Phi^{N,M}_{\prec \lambda}$. The quadratic term in $\mathcal{H}_{\min}$ is a first order differential operator and trivially satisfies the Leibniz rule. We thus focus on the quartic piece
\begin{align}
    \mathcal{H}^{(4)}_{\min}&(\Psi_{c,p}(z) \phi_{r(\lambda)}) = \sum^{i+j=k+l}_{i,j,k,l\ge 0} \min(i,j,k,l) z_i z_j \del_k \del_l \left( \Psi_{c,p}(z)  \phi_{r(\lambda)} \right) \nonumber \\
    &\hspace{-1.5em} =   \mathcal{H}^{(4)}_{\min}(\Psi_{c,p}(z))\phi_{r(\lambda)}  +  \Psi_{c,p}(z)  \mathcal{H}^{(4)}_{\min}(\phi_{r(\lambda)} ) + 2 \hspace{-0.5em}\sum^{i+j=k+l}_{i,j,k,l\ge 0} \hspace{-0.5em} \min(i,j,k,l) c p^l \Psi_{c,p}(z) z_i z_j  \del_k\phi_{r(\lambda)}  \, ,
    \label{eq: action Hmin on f Psi}
\end{align}
and show that the contribution from the third term is of lower order. We start by observing
\begin{align}\label{eq: Hmin monomial c2}
\sum^{i+j=k+l}_{i,j,k,l\ge 0}  p^l \min(i,j,k,l)z_i z_j \del_k \phi_{r(\lambda)}  &= \sum^{i+j\geq k}_{i,j,k\ge 0} p^{i+j-k} \min(i,j,k,i+j-k) z_i z_j \del_k \phi_{r(\lambda)} \, ,
\end{align}
and the goal is to invert the summation bounds, as we did in \eqref{eq: HGG c1}. The first step is to replace the $\min$ coefficient with $\max(0,\min(i,j,k,i+j-k))$, which removes the constraint on the indices $i+j\geq k$. Next, we use the identity
\begin{equation}
    \max(0,\min(i,j,k,i+j-k)) = \min(i,k)+\min(j,k)-k+(k-i-j)^+ \,
    \label{eq: min split}
\end{equation}
with $(x)^+=x$ if the argument is positive and zero otherwise. This trick allows for interpreting the infinite sums over $i$ and $j$ as lower-weight contributions plus a finite sum at each value of $k$ with bound $i+j<k$, which is also of a lower weight. We note that the first two contributions in \eqref{eq: min split} can be written as 
\begin{align}
    \sum_{i,j,k\ge 0} p^{i+j-k}\min(i,k) z_i z_j \del_k \phi_{r(\lambda)}  &= \sum_{j,k\geq 0} p^{j-k} \left( \sum_{i\geq 0}   p^{i} k z_i z_j \del_k  - \sum_{i= 0}^{k-1}  p^{i} (k-i) z_i z_j \del_k \right)\phi_{r(\lambda)} \,.
    \label{eq: Hmin c2 intermediate}
\end{align}
The first term yields a contribution of lower order since it has support in blocks $(N',M')$ of lower lexicographic order than $(N_1,M_1)$ by the same ensuing argument as the first term in \eqref{eq: HGG c1} for $\mathcal H$. For the second term in \eqref{eq: Hmin c2 intermediate} we use the identity
\begin{equation}
    c\Psi_{c,p}(z) \sum_{j \ge 0} p^{j} z_j  = \mathcal N \Psi_{c,p}(z) \, ,
    \label{eq: N on monomial app B}
\end{equation}
which follows immediately from the definition \eqref{eq: NM bargmann}. This shows that the second contribution to \eqref{eq: Hmin c2 intermediate} is also of lower order since $i < k$. 
Finally, the terms 
\begin{align}
   \sum_{i,j,k\ge 0} p^{i+j-k} (k-i-j)^+ z_i z_j \del_k \phi_{r(\lambda)} = \sum^{i+j< k}_{i,j,k\ge 0} p^{i+j-k} (k-i-j) z_i z_j \del_k \phi_{r(\lambda)} 
\end{align}
originating from the last contribution to \eqref{eq: min split} are combinations of states \eqref{eq: basis products psimax} of strictly lower weight $M'$ compared to $\phi_{r(\lambda)}$ since $k-i-j>0$. Multiplied with $\Psi_{c,p}(z)$ and projected to the block $(N,M)$, this yields a contribution of strictly lower lexicographic order \eqref{eq: ordering phis} to $\phi_\lambda$. 
This demonstrates the approximate Leibniz rule \eqref{eq: selfsimilarity hmin} for $\mathcal{H}_{\min}$.

\subsection{Upper triangularity from recursively applying the Leibniz rule}

We now demonstrate how the approximate Leibniz rules \eqref{eq: selfsimilarity hgg} and \eqref{eq: selfsimilarity hmin} imply an upper-triangular structure for $\mathcal H$ and $\mathcal H_{\min}$ in the ordered basis \eqref{eq: basis for ordering}. We consider \eqref{eq: selfsimilarity hgg} in detail, but the same argument holds for $\mathcal H_{\min}$. We start by noting that the first term in the Leibniz rule clearly contributes to the diagonal of the matrix representation of $\mathcal H$. The goal is to show that by recursively applying the Leibniz rule to the second term in \eqref{eq: selfsimilarity hgg}, where $\mathcal H$ acts on the lower block states $\phi_{r^{i}(\lambda)} = \psmax{n_{i+1}}{m_{i+1}} \phi_{r^{i+1}(\lambda)} $, one obtains an approximate Leibniz rule on the $d_\lambda+1$ components of the product \eqref{eq: basis products psimax final}:
\begin{align}  
    &\mathcal{H}(\phi_{\lambda}) 
    =    \sum_{j=1}^{d_\lambda+1} \mathcal{H} \left( \psmax{n_j}{m_j}  \right)  \prod_{i\neq j}^{d_\lambda+1} \psmax{n_i}{m_i}\, .
    \label{eq: full Leibniz rule}
\end{align}
We can see how this works by applying the Leibniz rule again to $\phi_{r(\lambda)} = \psmax{n_{2}}{m_{2}} \phi_{r^{2}(\lambda)} $ in \eqref{eq: selfsimilarity hgg}: 
\begin{align}  
    &\mathcal{H}(\phi_{r(\lambda)}) 
    =  \mathcal{H}(\psmax{n_2}{m_2})  \phi_{r^2(\lambda)}  + \psmax{n_2}{m_2} \mathcal{H}( \phi_{r^2(\lambda)})   \,
\end{align}
up to terms in $\Phi^{N_1,M_1}_{\prec {r(\lambda)}}$. This expression can then be inserted in \eqref{eq: selfsimilarity hgg} and the approximate Leibniz rule continues to hold if  $\psmax{n_1}{m_1} \Phi^{N_1,M_1}_{\prec {r(\lambda)}} \subseteq \Phi^{N,M}_{\prec {\lambda}}$. Indeed, such products are of the form \eqref{eq: basis products psimax}, manifestly of lower order \eqref{eq: ordering phis} compared to $\phi_{\lambda}$ and thus supported in $\Phi^{N,M}_{\prec {\lambda}}$. This step can be carried through recursively through all hook layers of $\lambda$ until one reaches $\phi_{r^{d_\lambda}(\lambda)}$, which is a single highest energy state and thus an eigenvector of $\mathcal{H}$. 
These steps demonstrate the validity of \eqref{eq: full Leibniz rule} and therefore the upper triangularity of the Hamiltonians \eqref{eq: HGG Bargmann} and \eqref{eq: Hmin Bargmann} in the basis \eqref{eq: basis products psimax final}. The approximate Leibniz rule \eqref{eq: full Leibniz rule} generalized to products of more than two highest energy states moreover provides a closed-form expression for the integer spectrum of the Hamiltonians in terms of sums of highest energy eigenvalues \eqref{eq: energies highest state}, as we shall proceed to state explicitly.

\section{The integer spectrum of $H$ and $H_{\min}$ and their joint eigenvectors} 
\label{sec: eigenbasis}

The upper-triangular structure of the Hermitian operators $\mathcal H$ and $\mathcal{H}_{\min}$ in the basis $B_{N,M}$ allows one to read off the spectrum of these operators from the diagonal elements of the respective matrices, given by \eqref{eq: full Leibniz rule}, and to derive a complete eigenbasis via a Gram-Schmidt process. The observed integer spectra admit an elegant closed form in terms of the hook lengths of the corresponding partitions.

We start with computing the eigenvalues by applying the approximate Leibniz rule on the states \eqref{eq: basis products psimax final}. 
The eigenvalue corresponding to the product-state is a sum of the eigenvalues of its individual factors, given by (\ref{eq: energies highest state}).
For $\mathcal{H}_{\min}$, one finds
\begin{equation}\label{eigHmin}
    E_{\min,\lambda} = \mathcal H_{{\min},\lambda\lambda} =\sum_{i=1}^{d_{\lambda}+1} e^{\min}_{\scriptscriptstyle{N_{i-1}-N_i,M_{i-1}-M_i}} = \sum_{i=1}^{d_{\lambda}} m_i^2=  \sum_{i=1} ^{d_{\lambda}} h_\lambda(i)^2  \, .
\end{equation}
The degeneracies of $\mathcal H_{\min}$ thus correspond to the multiplicity of partitions with the same principal hook length expansion, visualized in Fig.~\ref{tab: principal hooks}.

For $\mathcal{H}$, one obtains
\begin{align}
    E_{\lambda} = \mathcal  H_{\lambda\lambda} =\sum_{i=1}^{d_{\lambda}+1}e_{\scriptscriptstyle{N_{i-1}-N_i,M_{i-1}-M_i}} &= \frac12 \sum_{i=1}^{d_{\lambda}+1} \left( n_i -1\right) \left( 2m_i+n_i \right)
    =\frac12 \sum_{i=1}^{d_{\lambda}+1} \left( n_{i} -1\right) \left( 2h_\lambda(i)+n_i \right)  \nonumber
    \\
&= \frac12 \left(\sum_{i=1}^{d_{\lambda+1}} h_{\lambda+1}(i)^2 - \sum_{i=1}^{d_{\lambda}} h_{\lambda}(i)^2 -2M-N \right) \,,
    \label{eq: H eigenvalues 2}
\end{align}
where $\lambda+1 \in \mathcal{P}_{N}(M+N)$ is the partition obtained by adding one unit to all parts of $\lambda$ (including to the zero parts). The final expression \eqref{eq: H eigenvalues 2} for the eigenvalues of $\mathcal H$ encodes the identity \eqref{eq:shift joint eigenvectors} and follows from the observation that $\lambda +1$ has Frobenius coordinates 
\begin{align}
\lambda+1 &= (a_1+1,a_2+1,\dots,a_{d_{\lambda}}+1,0|N-1,b_1-1,\dots,b_{d_{\lambda}}-1) \nonumber \\
&= (a_1+1,a_2+1,\dots,a_{d_{\lambda}}+1,0|N_0-1,N_1-1,\dots,N_{d_{\lambda}}-1)
\end{align}
with $\{a_i,b_i\}$ the Frobenius coordinates of $\lambda$ and where $(a_{(d_{\lambda}+1)}(\lambda+1),b_{(d_{\lambda}+1)}(\lambda+1))$ exist only if $b_{d_{\lambda}} > 0$. Therefore, one has
\begin{equation}
     h_{\lambda+1}(i)^2 - h_{\lambda}(i)^2 = n_i(2h_\lambda(i)+n_i) 
     \label{eq: difference hooks}
\end{equation}
for $i = 1, \dots, d_{\lambda}$ and 
\begin{equation}
     \sum_{i=1}^{d_{\lambda}}(2h_\lambda(i)+n_i) = 2M +N - b_{d_{\lambda}}\, .
     \label{eq: 2M + N}
\end{equation}
When $b_{d_{\lambda}}>0$, the last term in \eqref{eq: 2M + N} together with the nontrivial contribution from $i=d_{\lambda}+1$ in the first line of \eqref{eq: H eigenvalues 2} cancels $h_{\lambda+1}(d_{\lambda+1})^2$, and \eqref{eq: difference hooks} combined with \eqref{eq: 2M + N} then prove the last equality in \eqref{eq: H eigenvalues 2}.


\section{Discussion}

We have presented a solution of the quantum Hamiltonians (\ref{eq: HGG}) and (\ref{eq: Hmin}), based on their Bargmann representations (\ref{eq: HGG Bargmann}-\ref{eq: Hmin Bargmann}). The solution relies on an explicit basis made of products of the previously known top eigenvectors in the Bargmann representation, as in (\ref{eq: basis products psimax}). An appropriate selection of such states, labelled by integer partitions, with an appropriate ordering, yields a basis in which both Hamiltonians are upper-triangular.
Their diagonal matrix elements then immediately give the integer eigenvalues (\ref{eigHmin}-\ref{eq: H eigenvalues 2}) expressed through the partition data,
while the eigenvectors may be recovered algorithmically via an ordinary Gram-Schmidt process, without any explicit matrix diagonalization.

The basis that plays a central role in our construction is intimately related to the quantum coherent states \eqref{eq: coherent multisoliton} corresponding to the
multisoliton configurations \eqref{classinv}  of the classical theory. When $R$ is finite in \eqref{eq: coherent multisoliton}, only states corresponding to partitions of Durfee size up to $R$ occur in the expansion of \eqref{eq: coherent multisoliton}, suggesting that the Durfee size of partitions is a good measure for the number of solitons present in the corresponding quantum eigenstates. We expect further systematic relations between the classical and quantum theories in our setting, in parallel to the related considerations for the Benjamin-Ono equation in \cite{Moll}.

In our treatment here, we have presented an elementary and self-contained solution for the spectra of the two Hamiltonians (\ref{eq: HGG}) and (\ref{eq: Hmin}), ignoring the presence of infinitely many other polynomials in the creation-annihilation operators that commute with them, providing quantum counterparts to the classical Lax structure and partly characterized in \cite{superintegrable}. These higher conservation laws are known empirically to possess integer spectra as well, and it is natural to expect that formulas for their eigenvalues will exist in terms of the partition data so as to complement our expressions (\ref{eigHmin}-\ref{eq: H eigenvalues 2}). Ideally, one would hope to obtain eigenvalue formulas
for the entire hierarchy of conservation laws in terms of their generating function(s), similar to the considerations for the Benjamin-Ono equation in \cite{qBO}, where partition data also play a central role.

While we know that the eigenvectors of  (\ref{eq: HGG}) and (\ref{eq: Hmin}) form a highly structured system of multivariate polynomials in the Bargmann representation,
and we have provided a simple Gram-Schmidt algorithm for recovering them,
we have not been able to identify them in terms of a pre-existing polynomial system, and that remains a subject for future work. In some of the eigenvector families,
natural explicit constructions utilizing the Hahn polynomials are possible, which we illustrate in appendix~\ref{app: hahn polynomials}. This will be explored further in \cite{OP}.

\section*{Acknowledgments}
We thank Sergey Derkachev, Hosho Katsura, Andrii Lyashik, Bal\'azs Pozsgay, Anton Pribytok, Ivan Sechin, Luc Vinet and Meri Zaimi for discussions.
MDC is supported by a Leverhulme Early Career Fellowship and benefited from the hospitality of the Perimeter Institute for Theoretical Physics through the
Emmy Noether Fellowship program during part of this project. Her work has furthermore been partly supported by the STFC consolidated grant ST/X000664/1. 
OE thanks Beijing Institute of Mathematical Sciences and Applications, and specifically Anton Dzhamay, for hospitality in March 2026, which provided space
for many interesting conversations related to the topics of this article. 
Our work has benefited from interactions with the generative AI system Claude (Anthropic) in exploring properties of the model and obtaining preliminary versions of some mathematical arguments.
\vspace{5mm}

\appendix
\numberwithin{equation}{section}
 
\section{Decomposition of product-states not arising from\\ reduced partitions}  \label{app: support}

We set out to prove the statement that the combination
\begin{equation}
    E_{n,m}^{N,M}=\psmax{N-n}{M-m}\psmax{n}{m} \in \mathcal{L}_{N,M},
\end{equation}
with the index assignment chosen among the two $\psi$-factors such that $(M-m,N-n) \geq _{\text{lex}}(m,n)$, lies in 
 \begin{align}
    V^{<_{\text{lex}}(m,n)} \equiv  \operatorname{span}\{  E_{n',m'}^{N,M}  :  (m',n') <_{\text{lex}}(m,n) \}
\end{align} 
when $2m+n \geq M$ or equivalently $m+n \geq M-m $. (The proof is formulated for $M>0 $. When $M=0$, the block is one-dimensional and the statement holds trivially.) The idea is to show that if a polynomial $f(z) \in \mathcal{L}_{N,M}$ lies in the orthogonal complement of $V^{<_{\text{lex}}(m,n)}$, then the condition  $2m+n \geq M$ necessarily implies that also $\langle f| E_{n,m}^{N,M} \rangle = 0$, using the Bargmann inner product (\ref{Bprod}), and hence $E_{n,m}^{N,M} \in V^{<_{\text{lex}}(m,n)}$.

Define the following generating function for the inner products $\langle f |E_{n,m}^{N,M} \rangle$:
\begin{equation}
\mathcal{G}_{f,12}(c_1,c_2,p_1,p_2) \equiv  \Big\langle f \Big |  \prod_{k=0}^\infty e^{(c_1 p_1^k+c_2p_2^k) z_k}   \Big\rangle = \hspace{-0.5em}\sum_{ (m,n) \leq_{\text{lex}} (M,N) } \hspace{-0.5em}   \langle f|E_{n,m}^{N,M}  \rangle \;    c_1^{N-n}c_2^n p_1^{M-m}p_2^m \,,
\label{eq: two soliton generating function for inner product}
\end{equation}
for a polynomial $f(z) \in \mathcal{L}_{N,M}$. The last representation follows from (\ref{eq: coherent psi max}). The sum only runs over parameters $(n,m)$ for which $E_{n,m}^{N,M}$ is defined, that is $0<n<N$ or $(n,m) \in \{(0,0),(N,M)\}$. For the purposes of the proof, it will be enough to consider 
\begin{equation}
    \mathcal{G}_f(c,p) \equiv \mathcal{G}_{f,12}(1-c,c,1,p) =  \hspace{-0.5em}\sum_{(m,n) \leq_{\text{lex}} (M,N) } \hspace{-0.5em}   \langle f|  E_{n,m}^{N,M}  \rangle  \;    (1-c)^{N-n}c^n p^m \,.
    \label{eq: generating function for inner products}
\end{equation}

The strategy is to show that $\langle f| E_{k,l}^{N,M}   \rangle = 0$ for all $(l,k) <_{\text{lex}}(m,n) $ implies $\langle f | E_{n,m}^{N,M}   \rangle = 0$ when $2m+n\geq M $, using the analytic properties of the function $\mathcal{G}_f(c,p)$. One useful property to note is that the contribution to $\mathcal{G}_f(c,p)$ of degree $c^d$ is divisible by $(p-1)^d$. This is most easily observed by viewing the coherent state exponent as $(1-c)+cp^{k} = 1+c(p^{k}-1) $ so that, expanding the exponential, every factor of $c$ comes with a polynomial in $p$ that is divisible by $p-1$. (If $k=0$, the combination $p^k -1$ vanishes.) 
Furthermore, $\mathcal{G}_f(c,p)$ respects the following inversion symmetry manifested by interchanging $(n,m) \leftrightarrow (N-n,M-m)$ in the sums:
\begin{align}
    \mathcal{G}_f(c,p) &= \hspace{-0.5em}\sum_{(m,n) \leq_{\text{lex}} (M,N) } \hspace{-0.5em}   \langle f| E_{n,m}^{N,M}   \rangle  (1-c)^{N-n}c^{n}p^{m} \nonumber \\&= \hspace{-0.5em}\sum_{(m,n) \leq_{\text{lex}} (M,N) } \hspace{-0.5em}   \langle  f | E_{N-n,M-m}^{N,M}  \rangle  (1-c)^{n}c^{N-n}p^{M-m} = p^M\mathcal{G}_f(1-c,1/p) \, ,
    \label{eq: scaling symmetry Gf}
\end{align}
where we used that the sum over $(m,n) \leq_{\text{lex}} (M,N) $ can be traded for a sum over $(M-m,N-n) \leq_{\text{lex}} (M,N) $ in the second equality, and $E_{n,m}=E_{N-n,M-m}$ in the last equality. The identity \eqref{eq: scaling symmetry Gf} implies the following relation. Consider the finite expansion of the generating function around $c=0$ as $\mathcal{G}_f(c,p) = \sum_{i} c^i P_i(p)$, and call $d_c$ the degree of the resulting polynomial for the variable $c$. The coefficient of the leading contribution $P_{d_c}(p)$ is a polynomial in $p$, with lowest and highest nonzero powers $d_{p_1}$ and $d_{p_2}$, respectively. The identity \eqref{eq: scaling symmetry Gf} then implies the relation 
\begin{equation}
    M = d_{p_1} + d_{p_2} \, ,
    \label{eq: constraint 1}
\end{equation} since $\sum_{i} c^i P_i(p) = \sum_{i} (1-c)^i P_i(1/p) p^M$ with the leading term in $c$ being equal on the two sides after expanding the r.h.s.

We now assume $2m+n\geq M $ and $\langle f|E_{k,l}^{N,M}   \rangle  = 0$ for all $(l,k) <_{\text{lex}}(m,n) $. An immediate implication of this assumption is that the contribution of degree $p^m$ to $\mathcal{G}_f(c,p)$ has coefficient 
\begin{equation}
    \mathcal{G}_f(c,p)\Big\vert_{p^m} =   \sum_{k\geq n}  \langle f| E_{k,m}^{N,M}   \rangle  \;    (1-c)^{N-k}c^k \,,
    \label{eq: contributionPm}
\end{equation} 
which can be read off from the definition \eqref{eq: generating function for inner products}. Second, $p^m$ divides $\mathcal{G}_f(c,p)$, and thus, in particular, $p^m$ also divides $P_{d_c}(p)$. Therefore, 
\begin{equation}
m \leq d_{p_1} \, .
\label{eq: constraint 2}
\end{equation}
In addition, since $P_{d_c}(p)$ is the polynomial multiplying $c^{d_c}$ in $\mathcal G_f(c,p)$ we also know that $P_{d_c}$ contains a factor $(1-p)^{d_c}$. Both functions $(1-p)^{d_c}$ and $p^m$ thus divide $P_{d_c}$, such that  
\begin{equation}
    d_c+m \leq d_{p_2} = M-d_{p_1} \leq M-m  \, ,
    \label{eq: constraint 3}
\end{equation}
where we used \eqref{eq: constraint 1} in the second step and \eqref{eq: constraint 2} for the last inequality. This last relation puts an upper bound on the degree of $\mathcal{G}_f(c,p)$ in $c$, which is found to be at most $M-2m$.

With the above ingredients, the aim is to show that the inner product $\langle f| E_{n,m}^{N,M}   \rangle$, i.~e.~the coefficient of the lowest degree contribution $c^n$ in \eqref{eq: contributionPm}, vanishes. From the starting assumption $2m+n\geq M $, we identify two cases. Either $2m+n> M $ or $2m+n= M $. In the first case, using \eqref{eq: constraint 3} we find 
\begin{equation}
    d_c \leq M-2 m< n \,.
\end{equation} But $n$ is the smallest allowed degree in \eqref{eq: contributionPm} for the variable $c$ while $d_c$ is the highest degree of the entire generating function $\mathcal{G}_f(c,p)$. Thus \eqref{eq: contributionPm} has to vanish identically to avoid a contradiction and, in particular, the coefficient of $c^n$ in \eqref{eq: contributionPm} vanishes and implies $\langle f| E_{n,m}^{N,M}   \rangle = 0$ as required. In the second case, by the same argument  $d_c \leq n$ so that \eqref{eq: contributionPm} is given by a single term $\mathcal{G}_f(c,p)\vert_{p^m} = \langle f|E_{n,m}^{N,M}   \rangle c^n$. However, using the defining relations \eqref{eq: two soliton generating function for inner product}-\eqref{eq: generating function for inner products}, we obtain
\begin{equation}
    \mathcal{G}_f(c=1,p)= p^M \langle f| \psmax{N}{M}   \rangle \, ,
\end{equation} 
with all contributions from lower powers of $p$ vanishing at $c=1$. This implies $\mathcal{G}_f(c,p)\vert_{p^m} = 0$ (since $m<M$, as $2m=M-n\leq M$) and thus $\langle f|  E_{n,m}^{N,M}    \rangle = 0$. This completes the argument in the case $2m+n=M$.

Thus, we found that $\langle f| E_{k,l}^{N,M}  \rangle = 0$ for all $(l,k) <_{\text{lex}}(m,n) $ implies $\langle f| E_{n,m}^{N,M}   \rangle = 0$ when $2m+n\geq M $. All functions $f(z)$ orthogonal to the span $V^{<_{\text{lex}}(m,n)}$ are therefore also orthogonal to $ E_{n,m}^{N,M} $ so that $ E_{n,m}^{N,M}  \in V^{<_{\text{lex}}(m,n)} $.

\section{Joint eigenfunctions of $\mathcal{H}$ and $\mathcal{H}_{\min}$ as  multivariate\\ Hahn polynomials}
\label{app: hahn polynomials}
We found that the eigensolution to the diagonalization of $\mathcal H$ and $\mathcal H_{\min}$ can be obtained in each block by applying the Gram-Schmidt procedure to the states \eqref{eq: basis products psimax final}, inspired by multi-soliton coherent states \eqref{eq: coherent multisoliton}. We have found that these eigenvectors have a formulation in terms of special multivariate orthogonal polynomials acting on the highest energy state in each block, by translating the closed form expression for the two families of eigenvectors derived in \cite{superintegrable} in the language of the present paper.
The occupation numbers of the individual modes are given in the Bargmann representation by the following differential operators:
\begin{equation}
    \hat \eta_k = z_k \del_{z_k}\,.
\end{equation}

The first family of eigenstates derived in \cite{superintegrable}, referred to as the top $H_{\min}$ eigenbasis, can be obtained as the result of a Gram-Schmidt process applied to states \eqref{eq: basis products psimax} with partitions 
\begin{equation}
    \lambda_k = (M-k,1^k),
    \label{eq: partitions first family}
\end{equation}
where $\eta_0 = N-k-1$ and $k = 0,1,\dots, \min(N,M)-1$. These partitions all reduce to the same reduced Young diagram, which is the empty diagram. (This, of course, is consistent with the fact that they share the same (maximal) eigenvalue for $\mathcal{H}_{\min} = M^2$.)

The result of the Gram-Schmidt process on the ordered states \eqref{eq: basis products psimax final} with partitions \eqref{eq: partitions first family} is expressible in a compact form in terms of the univariate Hahn polynomials in the variable $\hat\eta_0$ acting on the coherent state \eqref{eq: coherent psi max}
\begin{equation}
   \psi_{(0^k)}  = Q_k(\hat{\eta}_0,-\mathcal N-1,-\mathcal M,\mathcal N-1) \, \Psi_{c,p}(z)  \,,
    \label{eq: first family}
\end{equation}
with the Hahn polynomial defined in terms of the hypergeometric function
\begin{equation}
    Q_k(x,\alpha,\beta,\tilde{N})  \sim {}_3F_2(\{-k,-x,k+\alpha+\beta+1\};\{\alpha+1,-\tilde{N}\};1) \,.
\end{equation}
We use a normalization where the denominator of the coefficient of the highest order term in $x$ is independent of the parameters. Projecting \eqref{eq: first family} down to the block $(N,M)$ then yields $\min(N,M)$ independent eigenvectors of $\mathcal H$ and $\mathcal H_{\min}$.

The specified subscript for the states \eqref{eq: first family} denotes the reduced partition $\mu=r(\lambda)$, with zero-part multiplicities written out explicitly, associated with \eqref{eq: partitions first family}. The eigenvalue formula \eqref{eq: H eigenvalues 2} recovers the square integer pattern $E^{(0^k)} = \left( \frac{N + M -2k}{2} \right)^2 + C_{\emptyset}(N,M)$ derived in \cite{superintegrable}, where we group the block-dependent constant in $C_{\emptyset}(N,M)$. 
One can also verify that the states $\psi_{(0^k)}$ with $k \geq \min(N,M)$ vanish identically.

The second eigenvector family derived in section~10 of \cite{superintegrable} can be written in terms of the following product of two Hahn polynomials
\begin{equation}
   \psi_{(0^k1)} = Q_k(\hat{\eta}_0,\alpha_0,\beta_0,\mathcal N-2)\,Q_1(\hat{\eta}_1,\alpha_1,\beta_1,\mathcal N-\hat\eta_0-1) \, \psmax{N}{M}
    \label{eq: second family}
\end{equation}
for $k=0,1,\cdots,\min(N-1,M-3)-1$, and parameters $\alpha_0=-N$, $\beta_0=-(M-3)$, $\alpha_1=-(N-\hat\eta_0)-1$ and $\beta_1=-(M-N+\hat\eta_0)$. These states can be verified to follow from the subsequent Gram-Schmidt process applied to the states \eqref{eq: basis products psimax final} with the partition label 
\begin{equation}
    \lambda_k = (M-2-k,2,1^k) \, ,
\end{equation}
with multiplicities $\eta_0 = N-k-2$ and which indeed follow \eqref{eq: partitions first family} in the ordering \eqref{eq: ordering phis}.
Their reduced Young diagram corresponds to a single box and their energies are given by $E^{(0^k1)}_k = \left( \frac{N + M - 4 -2k}{2} \right)^2 + C_{(1)}(N,M)$ and $E_{\min} = (M-1)^2+1$, with the block-dependent constant $C_{(1)}(N,M)$ written out explicitly in \cite{superintegrable}.

This relation between solutions to the GG Hamiltonian and special multivariate orthogonal polynomials will be further investigated in future work \cite{OP}.

\subsection{Square integer patterns of eigenvalues in the spectrum \eqref{eq: H eigenvalues intro} of $\mathcal H$}
We observed numerically in \cite{superintegrable} that the entire spectrum appeared to be organized according to square integer sequences within degenerate eigenspaces of $\mathcal H_{\min}$. With the closed formulas and partition-labelled eigenfamilies, we are now able to describe the origin of this square integer pattern.

Extrapolating the preceding discussion, we expect the eigenvalues \eqref{eq: H eigenvalues 2} to belong to the same square integer sequence if their reduced partitions $\mu$ have the same Young diagram. Consider a Young diagram with $m$  boxes that can consistently be reduced from partitions in $\mathcal{P}_N(M)$ and denote the corresponding partition of $m$ with $\mu^0$. (We use the superscript $0$ to stress that this partition does \textit{not} contain information about the multiplicities of the part zero by itself, since we are currently agnostic about which partition in $\mathcal{P}_N(M)$ the Young diagram $\mu^0$ is reduced from.) The partitions $\lambda_k \in \mathcal{P}_N(M)$ from which $\mu^0$ can be obtained by hook-reduction are listed as 
\begin{equation}
   \lambda_k =  (M-m-\ell(\mu^0)-k,\mu^0+1,1^k)
\end{equation}
with multiplicity $\eta_0 = N-\ell(\mu^0)-k-1$ and parameter $k = 0,1,\dots,\min(M-m-\ell(\mu^0) - \mu^0_1, N-\ell(\mu^0))-1$ labelling the different partitions with same reduced Young diagram, and where $\mu^0+1$ adds one unit to each of the $\ell(\mu^{0})$ parts of $\mu^0$. The partitions only differ in their value of $N_1$ (the number of particles in the block to which $\mu^0$ belongs as reduced from $\lambda_k$), which takes values $N_1 = \ell(\mu^0)+k$. Inserting this in \eqref{eq: H eigenvalues 2} yields the square integer pattern
\begin{equation}
    E_{\lambda_k} = \left( \frac{N + M - m - \mu^0_1 - 2 \ell(\mu^0)-2k}{2} \right)^2 + C_{\mu^0}(N,M)
\end{equation}
up to a block-dependent constant $C_{\mu^0}(N,M)$ independent of $k$.

\end{document}